\documentclass[%
 aip,
 apl,%
 amsmath,amssymb,
reprint,%
]{revtex4-1}

\usepackage{graphicx}
\usepackage{dcolumn}
\usepackage{bm}
\usepackage{units}
\usepackage{todonotes}
\usepackage{hyperref}
\usepackage{xcolor}
\hypersetup{
    colorlinks,
    linkcolor={red!50!black},
    citecolor={blue!50!black},
    urlcolor={blue!80!black}
}

\begin{document}


\title{Temporal Feedback Control of High-Intensity Laser Pulses to Optimize Ultrafast Heating of Atomic Clusters}

\newcommand{\JAI}{The John Adams Institute for Accelerator Science, Imperial College London, London, SW7 2AZ, UK}

\newcommand{\IC}{Blackett Laboratory, Imperial College London, London SW7 2AZ, UK}%

\newcommand{\TATA}{Tata Institute of Fundamental Research, Homi Bhabha Road, Colaba, India}

\newcommand{\TIFR}{TIFR Center for Interdisciplinary Sciences, Hyderabad 500107, India}%

\newcommand{\RRC}{Laser Plasma Section, Raja Ramanna Centre for Advanced Technology, Indore 452013}%

\newcommand{\GOLP}{GoLP/Instituto de Plasmas e Fus\~{a}o Nuclear, Instituto Superior T\'{e}cnico, U.L., Lisboa 1049-001, Portugal}

\newcommand{\CLF}{Central Laser Facility, STFC Rutherford Appleton Laboratory, Didcot OX11 0QX, UK}

\newcommand{\UCL}{Department of Physics and Astronomy, University College London, London WC1E 6BT, UK}

\newcommand{\LMU}{Fakult\"at f\"ur Physik, Ludwig-Maximilians-Universit\"at M\"unchen, D-85748 Garching, Germany}

\newcommand{\MPQ}{Max-Planck-Institut f\"ur Quantenoptik, Hans-Kopfermann-Str. 1, D-85748 Garching, Germany}

\newcommand{\DESY}{Deutsches Elektronen-Synchrotron DESY, Notkestr. 85, 22607 Hamburg, Germany}

\newcommand{\CI}{The Cockcroft Institute, Keckwick Lane, Daresbury, WA4 4AD, United Kingdom}

\newcommand{\LANCS}{Physics Department, Lancaster University, Lancaster LA1 4YB, United Kingdom}

\newcommand{\UMICH}{Center for Ultrafast Optical Science, University of Michigan, Ann Arbor, MI 48109-2099, USA}

\newcommand{\SUPA}{SUPA, Department of Physics, University of Strathclyde, Glasgow G4 0NG, UK}

\newcommand{\LLNL}{Lawrence Livermore National Laboratory (LLNL), P.O. Box 808, Livermore, California 94550, USA}

\newcommand{\DLS}{Diamond Light Source, Harwell Science and Innovation Campus, Fermi Avenue, Didcot OX11 0DE, UK}

\newcommand{\YORK}{York Plasma Institute, Department of Physics, University of York, York YO10 5DD, UK}

\newcommand{\ELI}{ELI Beamline, Institute of Physics of the ASCR, Na Slovance 2, Prague 182 21, Czech Republic}

\newcommand{\LUND}{Department of Physics, Lund University, P.O. Box 118, S-22100, Lund, Sweden}

\author{M. J. V. Streeter}
\affiliation{\CI}

\author{S. J. D. Dann}
\affiliation{\CI}

\author{J. D. E. Scott}%
\affiliation{\CI}

\author{C. D. Baird}
\affiliation{\YORK}

\author{C. D. Murphy}%
\affiliation{\YORK}

\author{S. Eardley}
\affiliation{\IC}

\author{R. A. Smith}%
\affiliation{\IC}

\author{S. Rozario}
\affiliation{\JAI}%
\author{J.-N. Gruse}
\affiliation{\JAI}%
\author{S. P. D. Mangles}
\affiliation{\JAI}%
\author{Z. Najmudin}%
\affiliation{\JAI}%

\author{S. Tata}
\affiliation{\TATA}
\author{M. Krishnamurthy}%
\affiliation{\TATA}

\author{S.V.~Rahul}%
\affiliation{\TIFR}

\author{D. Hazra}%
\affiliation{\RRC}

\author{P. Pourmoussavi}
\affiliation{\DESY}%
\author{J. Osterhoff}%
\affiliation{\DESY}%

\author{J. Hah}%
\affiliation{\UMICH}%

\author{N. Bourgeois}
\affiliation{\CLF}
\author{C. Thornton}
\affiliation{\CLF}
\author{C. D. Gregory}
\affiliation{\CLF}
\author{C. J. Hooker}
\affiliation{\CLF}
\author{O. Chekhlov}
\affiliation{\CLF}
\author{S. J. Hawkes}
\affiliation{\CLF}
\author{B. Parry}
\affiliation{\CLF}
\author{V. A. Marshall}
\affiliation{\CLF}
\author{Y. Tang}
\affiliation{\CLF}
\author{E. Springate}
\affiliation{\CLF}
\author{P. P. Rajeev}
\affiliation{\CLF}

\author{A. G. R. Thomas}%
\affiliation{\CI}%
\affiliation{\UMICH}%

\author{D. R. Symes}
\affiliation{\CLF}%


\date{\today}

\begin{abstract}
We describe how active feedback routines can be applied at limited repetition rate (\unit[5]{Hz}) to optimize high-power (\unit[$>10$]{TW}) laser interactions with clustered gases. Optimization of x-ray production from an argon cluster jet, using a genetic algorithm, approximately doubled the measured energy through temporal modification of the \unit[150]{mJ} driving laser pulse. This approach achieved an increased radiation yield through exploration of a multi-dimensional parameter space, without requiring detailed \emph{a priori} knowledge of the complex cluster dynamics. The optimized laser pulses exhibited a slow rising edge to the intensity profile, which enhanced the laser energy coupling into the cluster medium, compared to the optimally compressed FWHM pulse (\unit[40]{fs}).  Our work suggests that this technique can be more widely utilized for control of intense pulsed secondary radiation from petawatt-class laser systems. 
\end{abstract}

\maketitle


Petawatt lasers are now able to operate with pulse repetition rates of \unit[1]{Hz} \cite{Nakamura2017DiagnosticsLaser} and upcoming facilities using more efficient, lower thermal load diode-pumped solid state technology will increase this to \unit[10]{Hz} \cite{Rus2015ELI-Beamlines:Systems} or more. One of the major drivers for the increase in \textit{average} power of such high peak-power systems is to generate bright laser-plasma based secondary sources to provide user beamlines, similar to existing light-source facilities, or energetic particle beams for a range of applications. This move to a higher repetition rate opens the possibility to employ active feedback routines to optimize energy conversion into radiation or particle beams. Due to highly complex non-linear dynamics in intense laser-plasma interactions, the optimal laser pulse properties for generation of the secondary source can not easily be predicted, as this is both computationally demanding and requires a complete understanding of all the key physical processes involved. Also, optimization is a many-dimensional problem and so cannot readily be performed by scanning individual parameters.

Sophisticated feedback techniques, usually employing kHz repetition rate lasers operating at relatively low peak intensity, are well-established for coherent control of atomic and molecular processes \cite{Warren1993CoherentAlive}. Programmable elements in the laser system are used to tailor the spatial and temporal profile of the laser pulse at focus to optimize specific output parameters. One method is to use a genetic algorithm (GA) to select the most suitable profiles out of an initially random or pseudo-random set and, over a number of generations, the input parameters are \textit{evolved} to improve the chosen output property (referred to as the \textit{fitness function}). 
The great benefit of this approach is that it can achieve advantageous results without detailed knowledge of the interaction itself, and lead to new and unexpected results. 

Previous experiments have employed feedback loops to control high harmonic generation \cite{Bartels2000Shaped-pulseX-rays,Yoshitomi2004Phase-matchedAlgorithm}, cluster dynamics \cite{Zamith2004ControlFragmentation,Moore2005ControlExplosion,Truong2011SystematicallyStudies} and electron beam properties \cite{He2015CoherentDynamics} through temporal and spatial pulse shaping. These studies were performed with low energy pulses (\unit[$<20$]{mJ}) and, with the exception of He {\it et al.} \cite{He2015CoherentDynamics}, at relatively low intensities (\unit[$<10^{16}$]{Wcm$^{-2}$}). Scaling these techniques to higher energy lasers is challenging because of the limited repetition rate. Although feedback has been used to improve the performance of high power (\unit[0.1--1]{PW}) lasers \cite{Liu2014Adaptive-feedbackPulses,Kim2017StablePulses}, it has not been applied \textit{directly} to optimize the secondary sources produced. 

Here, we report an experiment which adopted temporal shaping to optimize laser-driven x-rays using a laser capable of delivering much higher energy pulses (\unit[$\sim 1$]{J}) than previously used for such feedback routines. This set-up gives access to higher energy secondary sources such as high-\textit{Z} $K_{\alpha}$ sources \cite{Issac2004UltraFields, Hayashi2011OL}, \unit[200]{MeV} electron beams \cite{Geddes2015CompactCharacterization} and directional hard x-ray sources (keV--MeV) \cite{Corde2013FemtosecondAccelerators} through betatron oscillations and inverse Compton scattering \cite{Albert2016PPCF}. Our successful implementation of the method may also serve as a proof-of-principle demonstration for \unit[10]{Hz} PW systems currently being commissioned. 

As a target medium, we used a gas of argon clusters that provided a test system for optimization techniques because the dynamics of the interaction are complicated and heavily influenced by pulse shape \cite{Fennel2010Laser-drivenDynamics}. Because of their localized solid density, clusters absorb intense laser light much more efficiently than isolated atoms creating multi-keV electron temperatures \cite{Shao1996PRL}. Ultrafast K-shell radiation is emitted on the timescale of the laser pulse by hot electrons creating inner-shell vacancies in the high density cluster core \cite{Dorchies2008ObservationInteraction,Chen2010IntenseClusters}. The clusters subsequently explode and merge forming a hot low density plasma that expands hydrodynamically,  emitting x-ray radiation as it cools and recombines on the few nanosecond timescale \cite{Ditmire1995StrongClusters}. 
Atomic ionization begins early in the pulse, initially through field ionization and then collisionally as the electron density rises. The charge state can easily reach Ar$^{8+}$ and may reach Ar$^{16+}$ or Ar$^{18+}$ in the polar regions through polarization enhancement of the field \cite{Jungreuthmayer2004MicroscopicFields,Fukuda2006StructureFields}. 

At moderate laser intensities (\unit[$\sim$$10^{16}$]{Wcm$^{-2}$}) the strength of the laser field is insufficient to overcome the restoring field of the ions so electrons remain bound, forming a quasi-neutral nanoplasma \cite{Ditmire1996InteractionClusters,Milchberg2001PlasmaInteraction}. Inside the cluster the laser field is shielded while the nanoplasma is super-critical. This suppresses energy transfer until expansion reduces the electron density to $n_{e}\sim n_c$ (where  $n_c = \omega^2 m_e \epsilon_0/e^2$ is the critical density for laser frequency $\omega$), at which point the nanoplasma moves resonantly with the laser frequency. Many experiments in this regime have shown an enhancement in the laser-cluster coupling with a longer pulse (\unit[100's]{fs}) or multiple pulse structure \cite{Zamith2004ControlFragmentation,Moore2005ControlExplosion,Truong2011SystematicallyStudies,Fennel2010Laser-drivenDynamics}.

At higher intensities the laser removes a significant proportion of the cluster electrons thus invalidating purely hydrodynamic models. The extracted electrons gain energy from the laser field during multiple passes through the shielded core of the cluster \cite{Jungreuthmayer2004MicroscopicFields,Taguchi2004ResonantLight,Breizman2005NonlinearMicroclusters,Mulser2005TwoClusters,Kundu2006NonlinearInteraction} and can reach a temperature of \unit[10's of]{keV}, higher than the ponderomotive energy and much hotter than the \unit[1--2]{keV} limit of collisional heating. This process allows high absorption to be maintained even with very short (\unit[$<100$]{fs}) laser pulses \cite{Chen2002MeasurementPulses}. We show here that the x-ray generation in the short-pulse high-intensity regime is highly sensitive to the pulse shape and interpret our results as an optimization of the resonant heating of the electron cloud through collisionless processes.

\begin{figure}[ht]
	\includegraphics[width=8.5 cm]{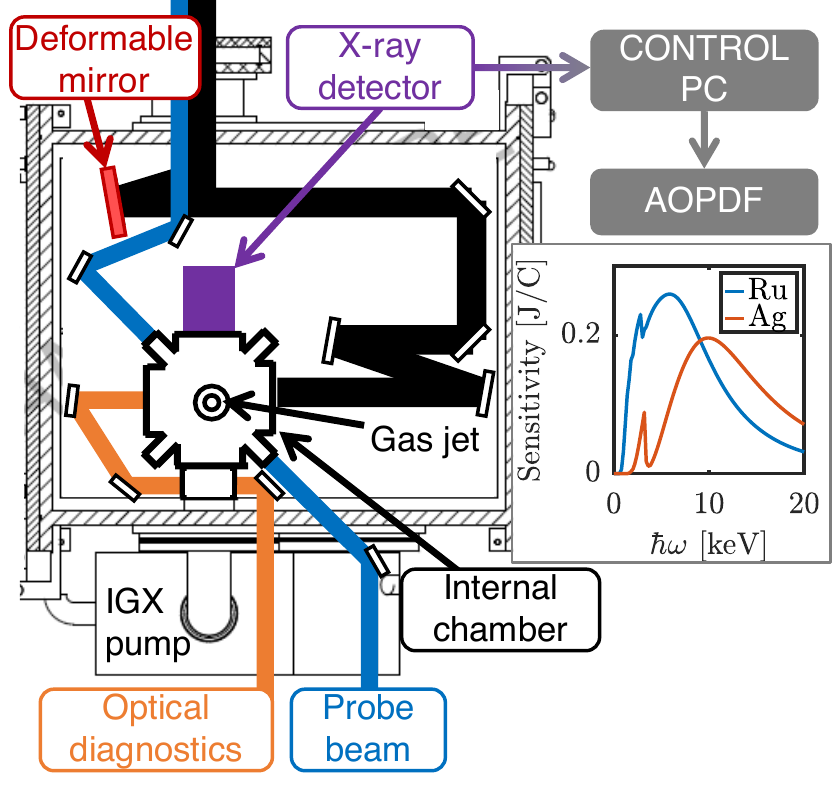}
	\caption{
		Experimental layout. The gas jet target is housed in an internal differentially pumped chamber. Diagnostic output is fed into the control PC that applies settings to the acousto-optic programmable dispersive filter in an optimization feedback loop. Sensitivity for the \unit[0.1]{$\mu$m} Ru and \unit[2]{$\mu$m} Ag filtered PIN diodes are shown.}
	\label{fig:FIG1}
\end{figure}

The experiment was conducted using the front end of the Gemini laser facility \cite{Hooker2006JDP}, which operates at \unit[5]{Hz} repetition rate. The arrangement is shown in Fig.~\ref{fig:FIG1}. The laser was focused with an \textit{f}/16 off-axis parabolic mirror to a spot size of \unit[22]{$\mu$m} FWHM with an on-target pulse energy of \unit[150]{mJ} and a pulse duration when fully compressed of \unit[40]{fs}. A deformable mirror was used to optimize the focal spot, giving a peak vacuum intensity of \unit[$\sim 4 \times 10^{17}$]{Wcm$^{-2}$}. An acousto-optic modulator (Fastlite \textit{Dazzler}) was used to modify the spectral phase of the laser pulse, and thereby modify the compressed pulse shape, which was diagnosed using frequency resolved optical gating (Swamp Optics \textit{Grenouille}). A \unit[3]{mm} diameter gas-jet produced argon clusters with an estimated radius \unit[${R}_{C} = 5\text{--}18$]{nm} over the \unit[7--40]{bar} ($1 \times 10^{18} - 5 \times 10^{18}$ atoms cm$^{-3}$) range of backing pressures \cite{Smith1998CharacterizationExperiments}. To reduce gas load in the main vacuum chamber, the jet was placed in an internal chamber with \unit[2]{mm} diameter entrance and exit holes to provide differential pumping. The laser energy before and after the compressor was monitored continuously to check for any drop in performance or degradation of the compressor throughput. X-ray generation was measured using two silicon PIN  diodes (Quantrad) mounted inside the internal chamber at \unit[90]{$^{\circ}$} to the laser axis, behind a \unit[50]{mm} \unit[0.09]{T} magnet to deflect electrons (\unit[$<100$]{keV}). One of these (model no. 025-PIN-125) was filtered with \unit[0.1]{$\mu$m} Ru and the other (model no. 100-PIN-250) with \unit[2]{$\mu$m} Ag. Both filters were held on \unit[3.5]{$\mu$m} mylar coated with \unit[0.1]{$\mu$m} Al. The spectral sensitivity of the diodes, calculated taking into account the filter transmission, is shown inset in Fig.~\ref{fig:FIG1}. Signals were averaged over 50 laser pulses to mitigate shot-to-shot fluctuations.

\begin{figure}[!htp]
	\includegraphics[width= 8.5 cm]{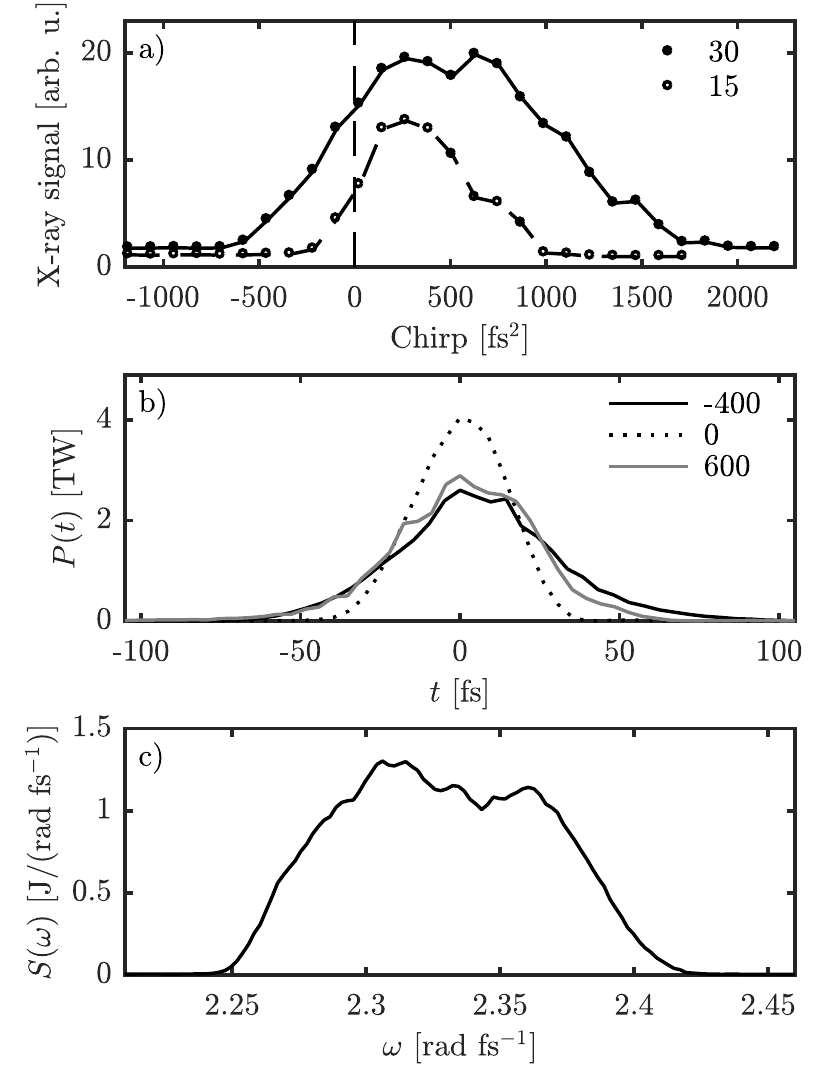}
	\caption{
		(\textbf{a}) Detected X-ray signal on a Ru-filtered PIN diode, plotted as a function of second order phase with backing pressure of \unit[15]{bar} and \unit[30]{bar}. (\textbf{b}) Laser pulse power profiles for second order phase terms of \unit[-400, 0 and 600]{fs$^2$}.  
(\textbf{c}) Laser spectrum.}
	\label{fig:FIG2}
\end{figure}

The effect on the x-ray signal of scanning the second order phase term (linear chirp) with the \textit{Dazzler} is shown in Fig.~\ref{fig:FIG2}(a) for two gas pressures, \unit[15]{bar} (\unit[$R_{C} \sim 8.5$]{nm}) and \unit[30]{bar} (\unit[$R_{C} \sim 14$]{nm}). Higher pressure ($P$) increases the overall atomic density linearly and also the cluster size \cite{Smith1998CharacterizationExperiments} approximately as $R_{C} \propto P^{3.8}$,  leading to stronger x-ray emission. In both cases the signal is not maximized at the position of shortest pulse duration (\unit[0]{fs$^2$} relative chirp), and shows a strong asymmetry with positive chirp yielding much higher signals than negative chirp. Example laser pulse profiles measured by the \textit{Grenouille} are shown in Fig.~\ref{fig:FIG2}(b). For a positive chirp (\unit[600]{fs$^2$}), an increased pulse duration results in an increased x-ray yield despite a drop in the peak power of the laser by a factor of 2. Even though a similar pulse shape is measured for a negative chirp (\unit[-400]{fs$^2$}), the x-ray yield is suppressed in this case. These results highlight the complex effects of spectral phase on x-ray yield, and perhaps the sensitivity to $\sim$100 fs timescale pulse contrast, which might also be affected by changing the \textit{Dazzler} pattern \cite{Liu2014Adaptive-feedbackPulses}. Including 3rd and 4th order spectral phase terms allows a much greater range of pulse shapes to be generated, including those with large asymmetries even with a symmetric spectrum as in this case (fig.~\ref{fig:FIG2}(c)). As the number of variables increases, however, performing scans of the high-dimensionality parameter space to find optimal conditions becomes unfeasible; instead, optimization algorithms such as genetic algorithms are more effective.

To implement the GA we defined the fitness function as the peak signal detected on the Ru-filtered PIN diode. This diode was used as it was sensitive to lower energy photons and therefore measured a detectable signal even when the x-ray yield was very low. A single \textit{generation} in the algorithm was formed from 15 \textit{individuals}, each individual being a set of spectral phase terms (2nd, 3rd and 4th terms of a polynomial expansion).  The initial generation always contained one individual with the unmodified settings (shortest pulse). The settings for the other individuals were randomly chosen from the permissible parameter ranges (limited to avoid damage to the laser chain). After evaluating the fitness function for each individual, the four best performing ones were selected to be the \textit{parents} for the next generation. Each \textit{child} individual in the new generation was created by a \textit{crossover} of two randomly selected parents. The crossover operation consisted of taking each phase term from one of the two parents at random. The children were further modified by \textit{mutation}, which consisted of adding random modifications to the phase terms, in order to maintain diversity and explore more of the permissible parameter space. In addition, the best performing individual from the last generation was always preserved and was the first one to be tested in the new generation. This allowed us to check that no significant change had occurred in the experimental conditions (larger than the normal level of fluctuation) during the time taken to acquire data for each generation (about 4 minutes). The feedback loop was continued until the fitness function appeared to converge to an optimum value.

\begin{figure}[!htp]
	\includegraphics[width=8.5 cm]{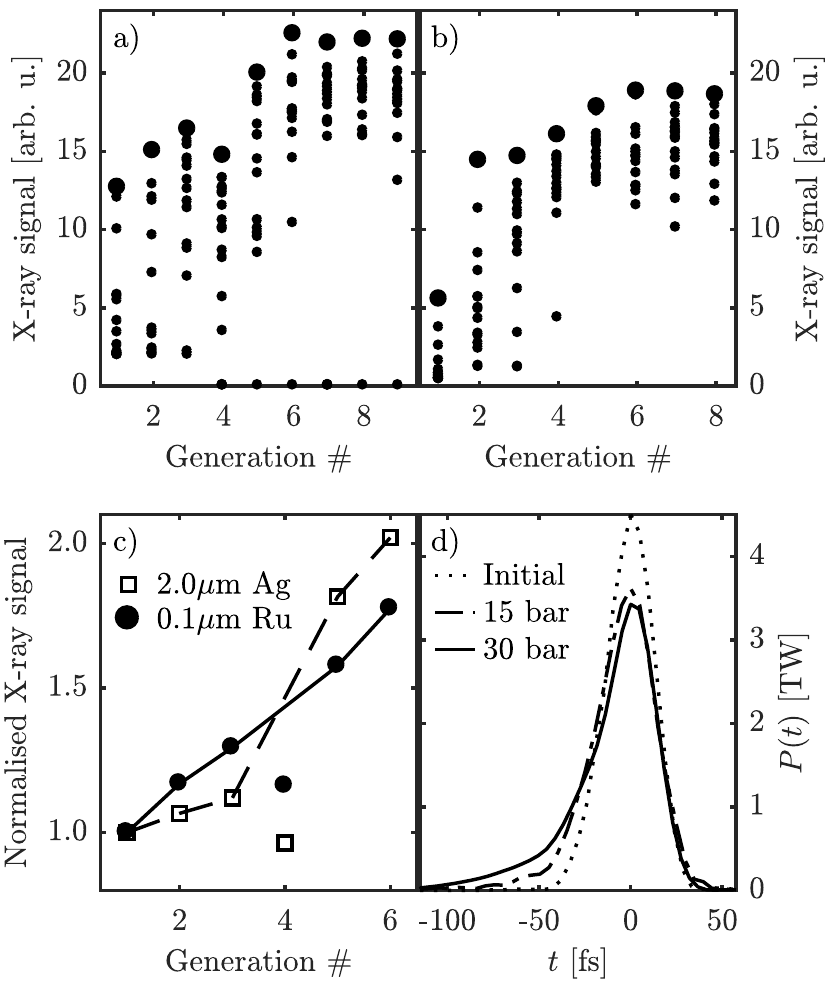}
	\caption{
		Optimization of Ru-filtered PIN diode X-ray flux with a backing pressure of (\textbf{a}) \unit[30]{bar} and (\textbf{b}) \unit[15]{bar}. Each point is the average of 50 shots, with the best individual of each generation shown as a larger point. Error bars are omitted for visual clarity.  (\textbf{c}) Improvement of X-ray signal through the \unit[0.1]{$\mu$m} Ru and \unit[2]{$\mu$m} Ag filters, normalized to their starting values with the unmodified laser pulse (Generation 1) for a backing pressure of \unit[30]{bar}. (\textbf{d}) Power profiles of the initial and optimized pulses from the \unit[15]{bar} and \unit[30]{bar} runs.}
	\label{fig:FIG3}
\end{figure}

An optimization of the signal over nine generations each containing fifteen members is shown in Fig.~\ref{fig:FIG3}(a) for \unit[30]{bar} backing pressure. The early generations show a wide variation because within a random choice of test patterns many generate a poor signal. The effect of `breeding' the best candidates becomes clear over later generations as the poor performers are rejected and the spread reduces. Generation 4 produced a low signal because of a laser defocus problem that was corrected before the start of generation 5. By generation 6, the x-ray flux reaches an optimum with $\sim$2$\times$ the value of the starting point and subsequent evolution does not further increase the signal. An equivalent scan for \unit[15]{bar} backing pressure shown in Fig.~\ref{fig:FIG3}(b) shows an improvement in x-ray flux of $\sim$3$\times$. In these cases we optimized only on peak voltage on the Ru-filtered PIN diode, but it is possible to define more sophisticated fitness functions such as signal ratios between Ross pair filtered diodes to increase plasma temperature. Even with our simple routine a higher increase in signal ($\sim2\times$) through the higher energy Ag filter suggests that the increased flux was accompanied by a rise in electron temperature (Fig.~\ref{fig:FIG3}(c)). The temporal profiles of the pre-interaction laser pulses at the end of each optimization run are shown in Fig.~\ref{fig:FIG3}(d)  compared to the starting point. In both cases the \textit{Dazzler} control signal evolved to apply 3rd and 4th order spectral phase terms as well as positive 2nd order phase. This acts to increase the asymmetry in the laser pulse, such that the rising edge is lengthened, while maintaining a relatively high peak power. 

Stronger x-ray emission indicates an increase in electron density or temperature and so is linked with more effective transfer of laser energy into the population of extracted electrons. Laser pulses with similar temporal shapes but opposite signs of second order phase (Fig.~\ref{fig:FIG2}(a)) differed in x-ray signal by a factor of 3.0 at \unit[30]{bar} and 5.3 at \unit[15]{bar}. This points to a dependence of ultrafast cluster dynamics that is more complicated than simply an optimum pulse duration and rather that the interaction is highly sensitive to subtle temporal profile changes on the \unit[10]{fs} timescale. This is not surprising since the collisionless heating of the extracted electron cloud has an extremely fast timescale and should be strongly affected by changes in the cycle-to-cycle structure of the laser field.

The most important consequence of pulse shaping is likely to be its effect on the expansion of the ionic core, the dynamics of which is primarily determined by the combined laser and electrostatic fields. With a sharp rising edge, ion motion is minimal and we estimate the fraction of liberated electrons as \unit[$\sim 15$]{\%} by equating the laser field with the binding field of the charged cluster (for \unit[$R_C = 14$]{nm}, $n_{e}=80 n_{c}$). A slower intensity rise triggers ionization earlier and the core has time to expand through thermal and Coulomb pressure over the course of the laser pulse. As the cluster expands, the charge density is reduced, making it easier for the laser to extract electrons. The cluster radius has only to increase by a factor of $\sqrt{2}$ (dropping the density to $\sim 30 n_{c}$) to double the number of extracted electrons. The preceding foot on the pulse is longer for higher gas-jet backing pressure since larger clusters expand more slowly. The temporal asymmetry seen in Fig.~\ref{fig:FIG2}(a) suggests that laser frequency chirp also  plays an important role in the rapidly evolving cluster. The system can be compared to a driven oscillator that is in resonance when the effective frequency of the electron cloud is matched to the laser frequency \cite{Mahalik2016AnharmonicStudy}. Over each cycle of rising intensity, the ionization and electron energies increase. It could be that a positive chirp (increasing laser frequency) maintains a resonant condition for many more cycles than in the opposite case. The exact combinations of spectral phase terms that were found by the GA would not have been easily reached by scanning each phase term individually and the finely tuned results demonstrate the advantage of using active feedback techniques. 

In summary, we have demonstrated the feasibility of applying active feedback control techniques with a \unit[$ > 10$]{TW} laser system operating at \unit[5]{Hz}. X-ray emission from an argon cluster plasma was optimized with a slowly rising intensity profile that improved the efficiency of the collisionless heating of electrons. Here we employed a genetic algorithm to optimize the signal from a single diagnostic, with three optimization parameters. However, there are many alternative algorithms available and one could use a larger parameter space and more intricate fitness functions to further improve experimental outcomes. Moreover, the availability of large-area fast-response piezo-electric based adaptive optics enables optimization routines to be applied also in the spatial domain. When combined with the acousto-optic modulator control of spectral phase, this provides the potential for complete control over the spatio-temporal properties of an intense laser focus to manipulate plasma interactions for specific desired outcomes. Our results suggest that the ability of the laser feedback system to control secondary sources is highly promising for future PW-class facilities planned for application-driven science. Employing this method on a dedicated beamline would provide the capability to initially optimize and then to continuously correct or tune the properties of an x-ray or particle source. 

We acknowledge funding from STFC grant numbers (ST/P002056/1 and ST/J002062/1) and the EuroNNAc, Newton-Bhabha and  Helmholtz ARD programs. S. E. is supported by an AWE CASE award. JH and AT acknowledges funding from the NSF grant 1535628 and DOE grant DE-SC0016804. We thank the staff of the Central Laser Facility for assistance with the experiment. 

%

\end{document}